# Ultrafast holography enabled by quantum interference of ultrashort electrons

**Authors:** I. Madan[1,5], G. M. Vanacore[1,5], E. Pomarico[1], G. Berruto[1], R. J. Lamb[2], D. McGrouther[2], T. T. A. Lummen[1], T. Latychevskaia[1], F. J. García de Abajo[3,4] and F. Carbone[1,*]

**Affiliations:**

[1]Institute of Physics, Laboratory for Ultrafast Microscopy and Electron Scattering (LUMES), École Polytechnique Fédérale de Lausanne, Station 6, CH-1015 Lausanne, Switzerland

[2]SUPA, School of Physics and Astronomy, University of Glasgow, Glasgow G12 8QQ, UK

[3]ICFO-Institut de Ciencies Fotoniques, The Barcelona Institute of Science and Technology, 08860 Castelldefels (Barcelona), Spain

[4]ICREA-Institucío Catalana de Recerca i Estudis Avançats, Passeig Lluís Companys 23, 08010 Barcelona, Spain

[5]Authors contributed equally.

[*]fabrizio.carbone@epfl.ch

**Abstract:**

Holography relies on the interference between a known reference and a signal of interest to reconstruct both the amplitude and phase of that signal. Commonly performed with photons and electrons, it finds numerous applications in imaging, cryptography and arts. With electrons, the extension of holography to the ultrafast time domain remains a challenge, although it would yield the highest possible combined spatio-temporal resolution. Here, we show that holograms of local electromagnetic fields can be obtained with combined attosecond/nanometer resolution in an ultrafast transmission electron microscope (UEM). Unlike conventional holography, where the signal and the reference are spatially separated and then recombined to interfere, in our method we use electromagnetic fields to split an electron wave function in a quantum coherent superposition of different energy states. In the image plane, spatial modulation of the electron-energy distribution reflects the phase relation between reference and signal fields, which we map via energy-filtered UEM. Beyond imaging applications, this approach allows implementing optically-controlled and spatially-resolved quantum measurements in parallel, providing an efficient and versatile tool for the exploration of electron quantum optics.

**Introduction**

Holography allows capturing both the phase and the amplitude of a signal distribution by superimposing it with a known reference. This method was originally proposed by Dennis Gabor to improve the resolution of an electron microscope. He first demonstrated the principle using light optics[1,2], while holography with electrons was shown shortly after[3]. With the invention of intense coherent light sources (lasers) and their most recent technological advancements, optical holography has become a popular technique for three-dimensional imaging of macroscopic objects, security applications[4,5] and microscopic imaging[6]. Electron holography[7,8] has been successfully employed in materials science[9], and also to image electrostatic potentials[10,11] and magnetic structures[12]. More generally, the holography principle can be extended to any kind of detection configuration involving a periodic signal undergoing interference, such as sound waves[13], X-rays[14], or femtosecond pulse waveforms (spectral holography)[15].

In recent years, various new experiments have been implemented to extend the concept of holography from a static imaging method to a dynamical probe, capable of recording the temporal evolution of both the amplitude and the phase of a signal. Time-resolved optical holography has been successfully realized in the femtosecond (fs) regime[16,17], and was used in combination with photoemission for plasmon imaging with enhanced spatial resolution in time-resolved photoemission electron microscopy[18,19] (tr-PEEM).

The introduction of temporal resolution in electron holography is more challenging, and so far the adopted schemes use temporal gating working in the microsecond timescale[20,21]. Reaching the ultrafast time-scales could become a reality owing to recent developments in UEM, in which femtosecond lasers are used to create ultrafast electron pulses[22,23]. This has enabled real-time filming of electronic collective modes[24–26], strain fields[27] and magnetic textures[28,29] with a temporal resolution down to a few hundred femtoseconds. Holography with ultrafast electron pulses should also be possible[23,30], providing similar time resolution and allowing subpicosecond imaging of magnetic and electric fields to be performed. However, most applications of interest involve spatially resolved phase-dynamics of electromagnetic fields on the timescale of one to a few femtoseconds. These include electronic excitations in condensed matter, non-radiative energy transfer in molecules, and excitonic currents and condensates, as well as optical fields in metamaterials and photonic crystals.

In this report, we demonstrate a time-domain holographic imaging technique implemented in UEM and based on the quantum coherent interaction of electron wavepackets with multiple optical fields. We illustrate this method by capturing attosecond/nanometer-resolved phase-sensitive movies of rapidly evolving local electromagnetic fields in plasmonic structures, as an example of the nanoscale imaging of phase dynamics. We implement two configurations of the experiment. The first and simplest one relies on using electron pulses to map the optical interference between a polaritonic reference and a polaritonic signal, both excited with ultrashort light pulses, similarly to near-field optical or tr-PEEM instruments. The second approach, which is conceptually different and unique to UEM, is based on a Ramsey-type interference[31], and relies on the coherent modulation of the electron wave function by means of *spatially separated* reference and signal electromagnetic fields. Since the interaction with the sequence of optical fields is occurring along the electron propagation direction, the constraint imposed by high transverse electron coherence necessary for conventional electron holography is now removed. This limitation has so far prevented the practical realization of holography in time-resolved experiments using pulsed electrons, especially when using multi-electron pulses necessary for imaging applications. Besides the obvious implications in the investigation of ultrafast coherent processes at the nanometer length scale, we show that our approach could be used for accessing the quantum coherence of generic electronic states in a parallel fashion, which can be relevant for electron quantum optics applications.

**Results**

*Conventional and holographic PINEM imaging*

A simple case of holographic UEM is based on *local* interference of two fields, which in the present study we realize with two propagating surface plasmon polaritons (SPPs). Next, we first describe the interaction mechanism of the electron pulse with a single SPP (Fig. 1a), and then examine the holograms produced by interference between the two SPPs (Fig. 1b).

In conventional Photon-Induced Near-field Electron Microscopy (PINEM)[24], electrons inelastically absorb or emit photon energy quanta $\hbar\omega$ (1.57 eV in our experiment). Filtering the inelastically scattered electrons allows one to form real-space images of the plasmon field[24]. The time resolution in PINEM is set by the duration $\tau_{el}$ of the electron pulses, which restrains the ability to perform real-space dynamical imaging to a few hundred fs[23–25]. Such time scale is in fact roughly two orders of magnitude longer than the plasmon period and even the duration of the plasmon wavepacket determined by the 55-fs light-pulse duration. While high spatial resolution can be achieved by monitoring plasmonic standing waves[32], for a travelling SPP the electron pulse duration limits the spatial resolution to $\Delta x \sim \tau_{el} v_g$, where $v_g$ is the plasmon group velocity. This is schematically shown in Fig. 1a. For the SPPs at the Ag/Si$_3$N$_4$ interface here studied, the resulting blurring in the real-space image is typically $\Delta x \sim 50$ μm, which is comparable with the plasmon decay length (~65 μm in our system).

This problem of blurring can be overcome *via* a holographic approach, which employs a second SPP wave, used as a reference, to create an interference pattern with the SPP of interest. Such interference forms *only when both waves overlap in space and time* (Fig. 1b). Denoting $\boldsymbol{k}_1$ and $\boldsymbol{k}_2$ the wave vectors of the two SPPs, the electric field associated with the interference pattern becomes

$$E_{\text{tot}}(\boldsymbol{r},t) = E\cos(\boldsymbol{k}_1\boldsymbol{r} - \omega t) + E\cos(\boldsymbol{k}_2\boldsymbol{r} - \omega t) = 2E\cos\left(\frac{(\boldsymbol{k}_1-\boldsymbol{k}_2)\boldsymbol{r}}{2}\right)\cos\left(\frac{(\boldsymbol{k}_1+\boldsymbol{k}_2)\boldsymbol{r}}{2} - \omega t\right). \quad (1)$$

Here, we consider only the component of the electric field that contributes to the interaction, namely the one parallel to the electron momentum[33]. As the electron interacts with this field over many optical periods, the resulting interaction has to be averaged over the time, thus reducing the contribution of the rightmost cosine factor of Eq. (1) to a constant factor in the inelastic intensity, eliminating spatial oscillations with $\boldsymbol{k}_1 + \boldsymbol{k}_2$ and leaving only those with $\boldsymbol{k}_1 - \boldsymbol{k}_2$. The resulting energy-filtered image is thus a hologram with interference fringes of period $2\pi/|\boldsymbol{k}_1 - \boldsymbol{k}_2|$ (see Fig. 1b).

When the SPPs are launched by independent pulses, one can tune the relative delay between the reference and the field of interest with sub-cycle precision (330 as in this work), therefore obtaining the real-space evolution of the phase of the electric field. Moreover, the finite duration of the reference pulse provides a temporal gate, effectively improving the temporal resolution of PINEM in tracking group velocities down to the duration of the laser pulse, in our case 55 fs. A similar concept has been presented in Ref. 34 to control the temporal profile of the electron pulse using a sequence of two incoherent interactions with a visible and near-IR pulses, whereas here the adoption of two phase-locked light pulses provides a fundamentally higher signal-to-background ratio (up to 100%) and gives direct access to the phase dynamics.

To demonstrate the holographic PINEM concept we have implemented the experiment described in Fig. 1b using a nanostructure composed of two perpendicular slits, fabricated by Ga-ion milling of a 43 nm thick Ag film deposited on a $Si_3N_4$ membrane (inset in Fig. 1). Each slit radiates SPPs at the interface between Ag and $Si_3N_4$ when excited with light polarized normally to its long edge (see Methods for more details). The experiment is conducted at a critical angle condition[35] that minimizes the interaction of the electron with the interrupted light beam (see Methods).

In Fig. 2 we show the holograms formed by the two SPPs with relative pulse delays of -77 fs, -20 fs, 0 fs, and 22 fs. These real-space images of the plasmonic field are obtained by energy-filtering inelastically scattered electrons[33,36] (see Methods). By varying the delay between the two light pulses, the position of such interference patterns changes across the square area delimited by the two slits, moving from the left-bottom part at negative times to the upper-right region for positive delays (see also Supplementary Movies 1 and 2). This demonstrates the gating effect of the reference pulse, showing that the envelope of the interference pattern is defined by the optical and not the electron pulse duration. The intensity profiles at each time delay plotted along the $\boldsymbol{k_1} - \boldsymbol{k_2}$ direction (marked by the arrow in In Fig. 2a) are shown in panels e-h.

Access to the phase dynamics allows us to measure the phase velocity $v_p$ (see Fig. 2i and Supporting Movie 2), while the improved temporal resolution of this method enables the determination of the group velocity $v_g$ (Fig. 2j and Supporting Movie 1). By taking into account the geometrical arrangements of the beams and the slits (see Methods and Supplementary Note I), we obtain $v_p$ = 2.69±0.05×10$^8$ m/s and $v_g$ =1.95±0.07 ×10$^8$ m/s, which agree well with the estimated theoretical values of $v_p$ = 2.64×10$^8$ m/s and $v_g$ =2.04×10$^8$ m/s, respectively. Because our technique is essentially a spatially-resolved temporal cross-correlation method, the characterization of the wavepackage cross-correlation can be done with arbitrary precision, in our case 330 as.

*Spatially-separated quantum holography*

The holographic approach presented above can be greatly generalized by utilizing the coherence between the different energy states of the quantum ladder in which the electronic wave function is split upon interacting with light[35,37,38]. This method exploits the fact that the electrons carry information about the amplitude and phase of the optical field even after the interaction is finished. Thus, the result of any ensuing interaction of the electron will depend on the relative phase between the initial and subsequent optical fields[31]. This allows us to separate in space the *interfering* fields, enabling in turn the adoption of more practical reference fields.

In particular, we make use of a semi-infinite light field created by the reflection of the optical beam from an electron-transparent optical mirror (Fig. 3a). The interrupted optical field interacts with the electrons via the inverse transition radiation effect[35,39], therefore creating a *material-independent* reference field, with nearly-constant spatial amplitude and phase, providing an optimum reference for holography.

The interaction with this first optical field is captured in the spatially homogeneous coupling factor $\beta_1$, which is a complex number uniquely determined by the amplitude and phase of the optical field, as well as the electron trajectory[33]. The interaction with the spatially varying signal field, occurring at a distance *d* further down the electron path, is captured by a space-dependent coupling factor $\beta_2(x,y)$. The total interaction is given by a simple sum of two complex numbers $\beta(x,y) = \beta_1 + \beta_2(x,y)$. The final energy distribution is defined by the modulus $|\beta(x,y)|$[33,40] which for slowly decaying plasmon fields is predominantly defined by the spatially dependent phase difference $\Delta\phi(x,y)$ between the two optical fields.

In Fig. 3b-d we show the experimentally measured variation of the electron energy distribution as a function of electron-beam distance from a directional plasmon emitter, formed as a result of the above-described spatially-separated interference. The distribution has the periodicity of the plasmon wavelength, both in the elastic and inelastic channels, which are in opposite phase (see Fig. 3d, where we plot the energy profile corresponding to the maximum (orange) and minimum (blue) of the elastic peak, which correspond to $\Delta\phi = \pi$ and 0, respectively).

The interaction strength $\beta$ depends on the electric field amplitude linearly as

$$\beta(x,y) = \frac{e}{\hbar\omega}\int_{-\infty}^{\infty} e^{-\frac{i\omega z}{v}} E_1(z) e^{-i\phi_1} dz + e^{-\frac{i\omega d}{v}} \frac{e}{\hbar\omega}\int_{-\infty}^{\infty} e^{-\frac{i\omega z}{v}} E_2(x,y,z) e^{-i\phi_2(x,y)} dz,$$

so the nonlocal interference contrast is mathematically equivalent to the previously discussed case of local interference. The only difference is a constant phase offset $e^{-i\omega d/v}$ between the two fields, which can be compensated by properly choosing the mutual delay between them. For fields that significantly vary not only within the $(x, y)$ plane but also along the $z$ coordinate (3D nano-objects, non-planar plasmonic structures, etc.), the phase-factor $e^{-i\omega d/v}$ produces an important contribution to the overall contrast. It reflects the change in the phase of the signal field due to retardation. In other words, it contains information on the $z$-distribution of the signal field, which can be directly retrieved because both the electron velocity and the light frequency are known quantities. This would allow for a complete three-dimensional (3D) phase tomography of the signal of interest to be performed, and could be used to reconstruct the complex electric field distribution around 3D particles or nanostructures.

The mathematical equivalence of local plasmon holography and spatially-separated quantum holography allows us to treat the recorded holograms with the same formalism of propagating and standing waves. In the spatially-separated case, the homogeneous reference field is non-collinear with respect to any propagating signal field, ensuring the formation of a standing wave pattern. An additional phase variation appears if the wavefront of the reference wave is tilted with respect to the mirrors surface, generating Doppler-like shifts[41] in the interference pattern.

We present an observation of this effect in Fig. 4. We record holograms formed by the tilted wavefront of the light reflected from a Ag mirror and a plasmon wave emitted from a hole carved in the Ag layer. The tilted wavefront can be described in Eq. (1) through the addition of the small in-plane momentum component $\boldsymbol{k_2} = \boldsymbol{k_{ph||}}$ coherently superimposing on the in-plane plasmon momentum $\boldsymbol{k_1} = \boldsymbol{k_{SPP}}$, whose direction is radial with respect to the center of the hole (see Fig. 4a). The resulting pattern exhibits a periodicity $\boldsymbol{k_1} - \boldsymbol{k_2} = \boldsymbol{k_{SPP}} - \boldsymbol{k_{ph||}}$, which is direction-dependent and leads to the Doppler effect shown in Fig. 4b,d. This effect is naturally absent in the untilted hologram shown in Fig. 4c.

*Detection of electronic quantum coherence*

Besides the direct implications in the visualization of phase-sensitive dynamics, our holographic approach can be useful for the characterization of the quantum state of a generic free-electron state, such as the one generated in the photoemission process from a solid-state material illuminated with UV light. This problem is of great interest not only for ultrafast TEM, but also for free-electron lasers and attosecond physics[42]. It has been shown that in the photoemission process electrons carry information about the phase of the exciting optical field[43]. However, due to several uncontrolled factors, such as screening potentials, scattering events, or external field inhomogeneities, coherence is usually lost in part[44]. While in attosecond science reliable techniques have been developed to investigate this issue[45,46], it is still a pending problem in UEM, which is particularly relevant when targeting sub-eV excitations[26] in condensed matter, where the relevant energy is smaller than or comparable with the electron energy spread, and thus beam coherence becomes an important condition.

In mathematical terms, quantum coherences of a state manifest in nonzero off-diagonal terms of its density matrix. In our method, the interaction between the SPP and the electron makes a generic electronic quantum state $\rho_{\text{in}}$ evolve unitarily into a space-dependent (distance $x$) state $\rho_{\text{out}}(x)$. In this way, off-diagonal terms of $\rho_{\text{in}}$ get projected onto the observable diagonal terms of $\rho_{\text{out}}(x)$. As our approach is able to simultaneously record spatial and spectral information (see Fig. 3b), we can readily determine how the energy distribution of the final electronic state varies with respect to $x$. This information can be used to identify and characterize the quantum coherences of the initial state.

The model calculation presented in Fig. 5 shows how our method can discriminate between a highly coherent (pure) and a fully incoherent (completely mixed) electron distribution, modeling the density matrix $\rho_{\text{in}}$ of photoelectrons generated for example by UV illumination of a solid target. These states are then made to interact with a travelling plasmon polariton excited by a mid-infrared optical field considered to be phase-locked with respect to the UV light and with a photon energy significantly smaller than the energy width of the photoemitted electrons. First, we consider electrons emitted in a pure Gaussian state (Fig. 5a,b, see Methods for details), for which the coherent interaction with the SPP field results in a generally asymmetric spectrogram, whose shape strongly depends on the phase difference between the SPP field and the UV photon used in the photoemission process. We stress that this phase dependence is a general property of any pure state spectrogram, while the asymmetry is not necessarily observed, and might be absent for some particular profiles of the wave functions. In contrast, when considering the spectrogram of a completely mixed Gaussian state (Fig. 5c,d), we find it to be symmetric and phase-independent. Thus, by observing the spatial dependence of the electron energy distribution, we can establish whether there is partial coherence in the photoemitted electrons.

This observation allows us to propose a further extension of UEM holographic imaging discussed above. Indeed Fig. 5b suggests that the spectrogram formed by coherent photoelectrons encodes information about the spatial phase distribution of SPPs, even without making use of the reference optical field, providing the most practical realization of quantum-holographic UEM.

Similarly to the approach described by Priebe *et al*.[47], our methods also provide sufficient information for the reconstruction of the complete density matrix of an unknown electronic state, but now, in contrast to their approach, consisting in obtaining the density matrix after multiple acquisitions with different phase or amplitude field configurations, our scheme uses well-controlled spatially dependent SPP fields to realize a number of projective measurements in a parallel fashion. Besides purely practical aspects, the intrinsically parallel nature of the acquisition method can be interesting in evaluating entanglement between single electrons. Indeed, the position-dependent unitary interaction not only allows us to project the electron state onto different measurement bases, which is a necessary tool for revealing entanglement, but it also allows us to do it for different electrons simultaneously. This can be achieved by using engineered apertures that direct entangled electrons through the desired regions of the optical field. Additionally, coincident detection of the spectrum of entangled electrons, directly applicable to our parallelized scheme, makes it possible to implement a Bell-type experiment for electrons.

**Conclusions**

In this work, we have demonstrated both local and spatially-separated holographic approaches based on ultrafast transmission electron microscopy. We have shown that these methods significantly improve the time resolution and, because they are phase sensitive, they allow us to determine the phase and group velocities of the propagating SPPs involved in the experiment. Moreover, the nonlocal character of our method allows us to completely decouple the reference and probe fields, which is not possible when relying on near-field-optical or photoemission microscopy techniques. We highlight that our demonstration of spatially-separated quantum holography is enabled by exploiting the interaction with a semi-infinite light field, which provides a nearly-perfect material-independent reference. The extension of this method to any local collective field promoting periodic modulation of the electron wave function is straightforward, and offers a unique perspective to achieve atomic and sub-femtosecond combined resolution in TEM. Possible objects of interest to be studied with this technique are atomic polarizabilities, excitons, phonons, Higgs and other collective and quasiparticle excitations in condensed matter systems. In addition, our method enables a spatially-resolved detection method of coherences in electron quantum states with great potential for electron quantum optics applications.

## Methods

*Experimental*

Electron-transparent samples were prepared by depositing through sputtering a silver thin film (43 nm) on a $Si_3N_4$ membrane (30 nm) suspended on a 80 × 80 µm² window of a Si support. Grooves completely penetrating the Ag film and partially penetrating the $Si_3N_4$ membrane were produced by Focused Ga-Ion Beam milling. The samples were mounted in a double-tilt TEM holder.

The 55-fs long, 1.57-eV, 300-kHz repetition-rate optical pulses from a Ti:Sapphire regenerative amplifier were used as both signal and reference field. The third harmonic of these pulses was used to produce electron pulses via photoemission from a $LaB_6$ cathode. Experiments are conducted in a modified JEOL-2100 TEM, described in Ref. 48.

Energy filtered imaging was performed using a Gatan Quantum GIF electron energy-loss spectrometer.

In our experimental setup, the light beam propagates at an angle of 4.5±1° with respect to the direction of the electron beam, which is oriented along the optical axis of the microscope. For the local interference experiment, the sample was tilted by 12° with respect to electron direction while maintaining the normal to the surface within the plane defined by the light and electron beam directions. This allows us to minimize the contribution from the semi-infinite field reflected by the flat sample surface, as for these conditions the electron wave function modification by the incident and reflected beams exactly cancell[35]. The presence of such tilt results in retardation effects, which are manifested as a tilt of the plasmon wavefront with respect to the edge of the groove sources. The images presented in Figs. 2 and 4 are projections of the electron distribution on the plane perpendicular to the optical axis of the microscope. Geometrical considerations accounting for the retardation effects and the calculation of the corresponding corrections to the phase and group velocities are presented in the Supplementary Note I.

*Data analysis*

The data presented in Figs. 2 and 4 were acquired in the ZLP-suppression mode described in Ref. 26. To make the contrast more visible, we subtracted a slowly varying background in Fig. 2, approximated by a 2D parabolic function and originating predominantly in the beam shape and a non-holographic PINEM offset. The magnification of the images was calibrated with a 463 nm optical replica calibration sample. Presented images were median-filtered to remove off-scale pixels produced by cosmic rays. The theoretical estimates of group and phase velocities have been conducted as described in Ref. 25.

*Calculation of electron spectra shown in Fig 3a*

Electron energy spectra shown in Fig. 3a has been calculated as described in Ref. 35, for electron and photon pulses of 60 fs, and $\theta = 35°$. Intensity of both reference and signal fields has been taken to be 0.13 GW/cm$^2$.

*Calculation of the spectrograms for the pure and mixed electronic states*

Two extreme types of electronic quantum states have been considered as input of the holographic technique considered in Fig. 5: a pure (p) Gaussian state with wavefunction $|\psi\rangle_p = \sum_i g(E_i) |E_i\rangle$, where $g(E_i)$ is a Gaussian probability amplitude, and a completely mixed (cm) Gaussian state with a diagonal density matrix $\rho_{cm} = \sum_i h(E_i) |E_i\rangle\langle E_i|$, where $h(E_i)$ is a Gaussian probability.

In the interaction picture, the two optical interactions implemented in the spatially-separated holographic method are described by the unitary operators[31,38,47,49]

$$U_1 = e^{\beta_1 a^\dagger - \beta_1^* a} \text{ and } U_2(x) = e^{\beta_2(x) a^\dagger - \beta_2(x)^* a},$$

where $\beta_1$ and $\beta_2(x)$ are complex coupling constants determined by the field amplitudes, while $a$ and $a^\dagger$ are commuting operators that lower and raise the number of electron-photon exchanges. Notice that $\beta_1$ is uniform over the sample plane, as it is determined by reflection of the normally incident light, whereas $\beta_2(x)$ shows a space dependence originating in the phase accumulated during plasmon propagation.

The initial states evolve into

$$\rho_{out}(x) = U_2(x)U_1\rho U_1^\dagger U_2(x)^\dagger = U_{tot}\rho U_{tot}^\dagger,$$

where $U_{tot} = U_2(x)U_1 = e^{\beta_{tot}a^\dagger - \beta_{tot}^* a}$ with $\beta_{tot} = \beta_1 + \beta_2(x)$.

The calculations in Fig. 5 are obtained by considering $\beta_1 = 0$ and $\beta_2(x) = A\,e^{i\Delta\phi}e^{ikx}$, where A and k are set to 5 and 1, respectively, and $\Delta\phi$ is the phase difference between the laser pulses used for photoemission and SPP generation. Absence of the first interaction allows us to demonstrate the possibility of performing holography with coherent electrons, without making use of the reference field.

## Acknowledgements

The authors thank dr. I. Kaminer (Faculty of Electrical Engineering and Solid State Institute, Technion, Haifa, Israel) and dr. M. Hassan (Department of Physics, University of Arizona, Tucson, USA) for fruitful discussions.

## Author contribution

T.T.A.L. designed the sample geometry; R.J.L. and D.M. fabricated the sample; I.M., G.M.V. and G.B. conducted the experiments; G.M.V. and I.M. analyzed the data; E.P. performed the calculations; I.M., E.P., G.M.V., F.J.G.d.A. and F.C. interpreted the results; I.M., E.P., G.M.V., G.B., T.L., F.J.G.d.A. and F.C. contributed to writing the article.

Authors claim no competing interests.

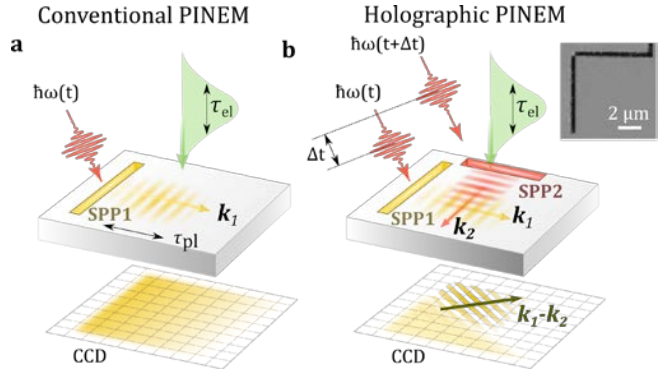

**Fig. 1. Conventional vs holographic PINEM imaging. a**, In conventional PINEM propagating, SPPs are imaged with long electron pulses, rendering only its time-averaged envelope with a spatial resolution $\Delta x \sim v_g \tau_{el}$. **b**, In local holographic PINEM, two SPPs propagate with orthogonal wave vectors $\boldsymbol{k}_1$ and $\boldsymbol{k}_2$, forming a standing-wave pattern along the direction $\boldsymbol{k}_1 - \boldsymbol{k}_2$, which is imaged as a periodic modulation in PINEM (the hologram). The interference contrast appears only when the two pulses overlap in space and time. The inset shows the SEM image of a fabricated structure. Black regions are grooves, which serve as plasmon sources.

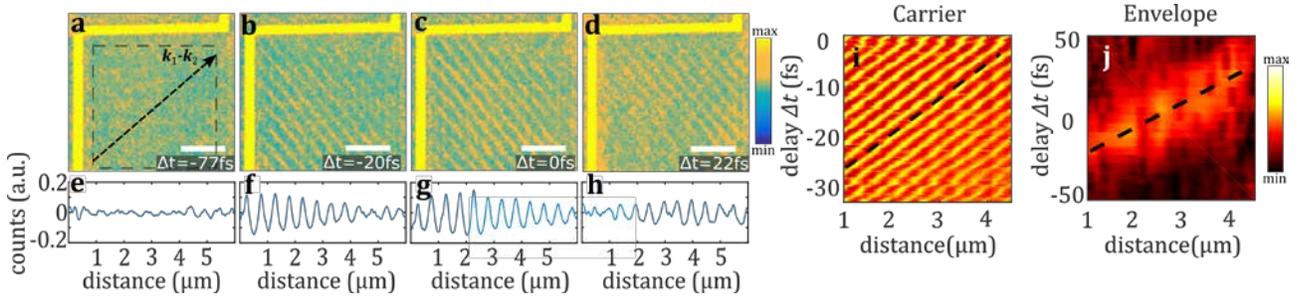

**Fig. 2. Holographic images formed by two pulses of orthogonal polarization at different delays. a-d**, Micrographs of PINEM images for different values of the relative time delay $\Delta t$ between the photoexciting pulses, as indicated in each image. The scale bars are 2 μm. The SPP emitted from the vertical slit propagates from left to right. Correspondingly, the interference pattern moves from the bottom-left to the top-right corner. **e-h**, Modulation of the electron counts along the $\boldsymbol{k}_1 - \boldsymbol{k}_2$ direction indicated in **a**, calculated as the average of counts along the direction orthogonal to $\boldsymbol{k}_1 - \boldsymbol{k}_2$, taken within the dashed square indicated in **a**. **i**, Evolution of the profiles shown in panels **e-h** as a function delay between the two pulses; because of the experimentally adopted sample orientation, retardation effects causes the slope of the fringes (dashed line as a guide) to be decreased by a factor of 0.71 with respect to the SPP phase velocity (see Methods). **j**, Envelope of the interference pattern as a function of delay between the two pulses, with the slope of the peak (dashed line as a guide)also decreased by a factor of 0.71 with respect to the SPP group velocity. Envelope data has been acquired in a separate measurement over longer delay span and with larger time steps.

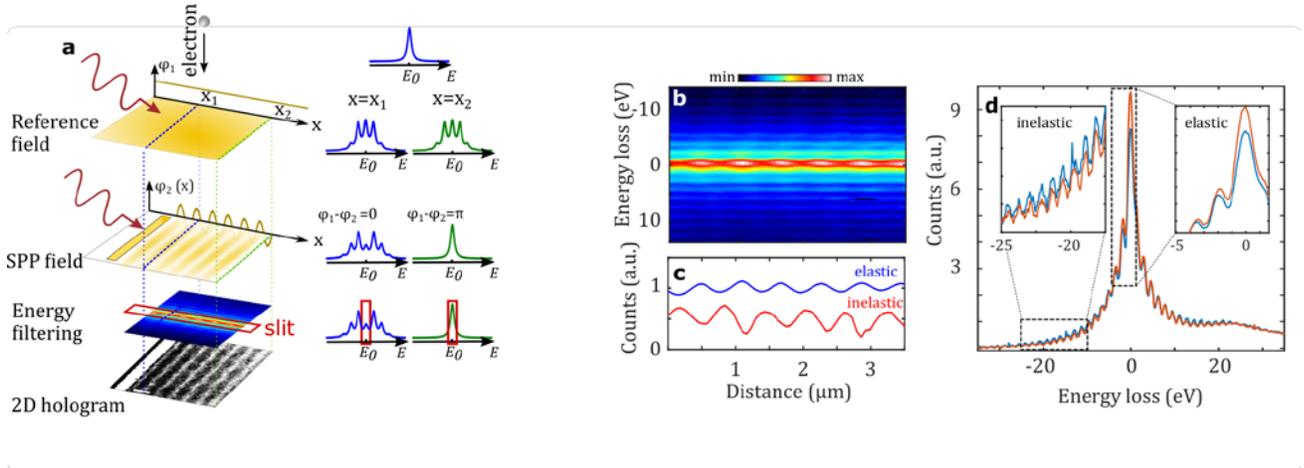

**Fig. 3. Principle of spatially-separated electron holography. a**, The initial energy distribution of the electron beam is a function of energy that is singly peaked at $E = E_0$ (on the right). Interaction with the reference field produces coherent superposition states with energies $E = E_0 \pm n\hbar\omega$. The ensuing interaction with an SPP depends on the relative phase between SPP and reference fields, which results in a position-dependent electron energy distribution. The elastic part of the electron spectrum is then used to form the 2D hologram. The spectra on the right are simulations from an analytical model (see Methods). **b**, Hybrid energy-space map (spectrogram) of the electrons after interaction with the two fields, as schematized in **a**. **c,** Spatial profiles of the normalized intensity for elastic (blue curve) and inelastic (orange curve) electrons, as obtained from panel **b** by energy-averaging from -1 to 1 eV for the elastic contribution and from -27 to -12 eV for the inelastic one. **d**, Energy profiles at the maximum and minimum of the spatial modulation shown in panel **b**, averaged over 4 periods.

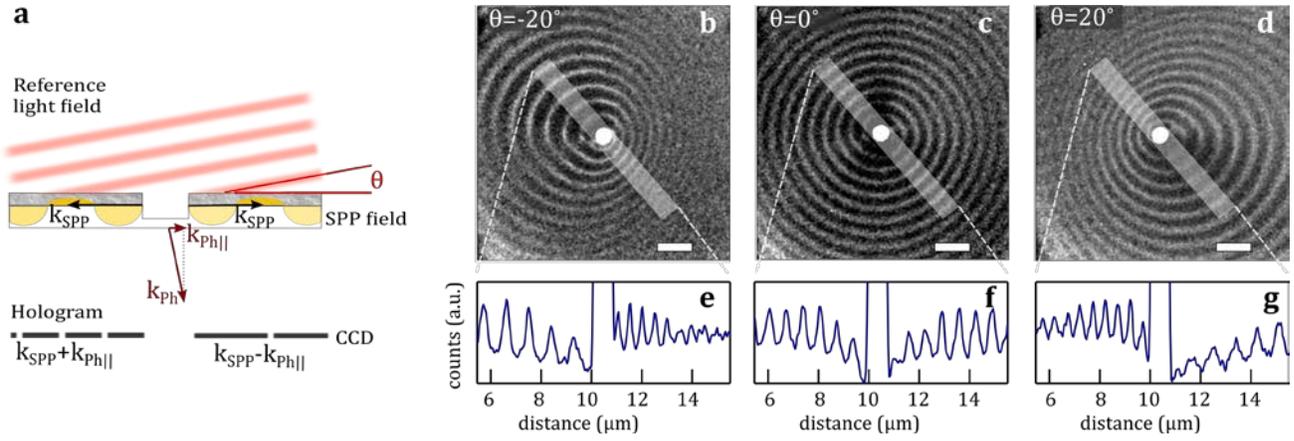

**Fig. 4. Doppler effect in spatially-separated electron holography. a**, Schematics of the experiment. Top: a laser pulse of frequency $\omega$ creates both a reference field and a signal field. The reference is produced by reflection from the surface, with an in-plane wave vector dependence on the tilt angle $\theta$ given by $k_{Ph||} = (\omega/c)\sin\theta$. The signal field is the SPP excited at the metal/dielectric interface, with a tilt-independent in-plane wave vector $k_{SPP}$. Bottom: change in the period of the hologram depending on the relative direction of $k_{Ph||}$ and $k_{SPP}$. **b-d**, Holograms observed for incident angles $\theta =$ -20°, 0°, and 20°, respectively. Scale bar is 2 µm. **e-g**, Spatially-resolved intensity profiles corresponding to the areas highlighted in white in **b-d**.

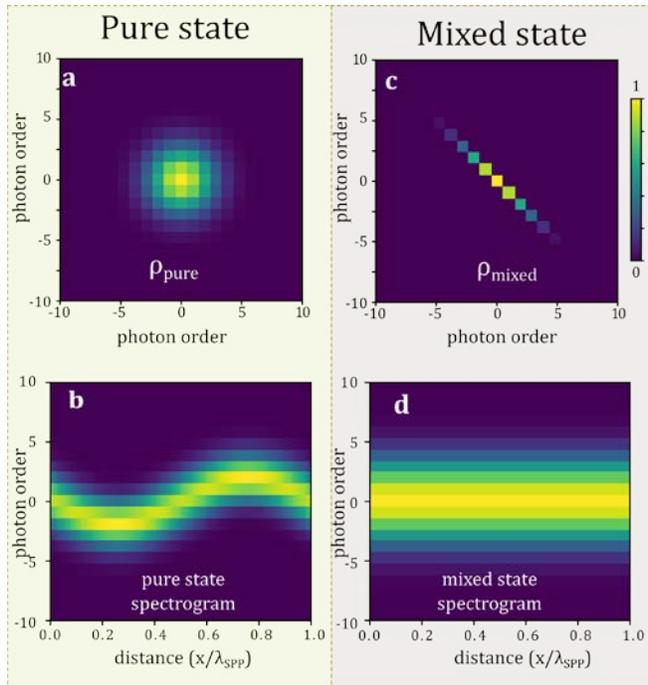

**Fig. 5. Proposal for the determination of the coherence of photoemitted electrons. a,** Density matrix of a fully coherent (pure) state created by photoemission. **b,** Spatially-dependent spectrogram formed after the interaction of the pure state with an SPP. **c,** Density matrix of the completely mixed state. **d,** Spectrogram formed after interaction of the mixed state with an SPP.